**COMPUTER MODELING OF EDGE EFFECTS IN PLASMA CRYSTALS.** J. A. Vasut and T. W. Hyde, Center for Astrophysics, Space Physics and Engineering Research, Baylor University, Waco, TX 76798-7316, USA, phone: 254-710-2511 (email: John_Vasut@Baylor.edu & Truell_Hyde@Baylor.edu).

**Introduction:** Dusty plasma systems play an important role both in astrophysical environments (protostellar clouds and ring systems) and laboratory situations (plasma processing).

Plasmas are generally regarded as a highly disordered state of matter. However, under certain conditions a plasma can exist in a crystalline state[1]. Such an ordered state will almost never occur (except perhaps in the interior of white dwarf stars) for a simple electron-ion plasma. However, it has been shown that the addition of dust grains as a component to the plasma makes such a crystallization possible[2]. This was first experimentally observed in 1994[3].

Such plasma crystals, also referred to as dust crystals or Coulomb crystals, are generally experimentally formed in a GEC RF reference cell where the bottom electrode provides the repulsion necessary to counteract the particle's weigh[4]. The crystals form in the plasma sheath between the electrode and the bulk of the plasma and usually take the form of a "two and a half" dimensional crystal. This peculiar situation is a result of gravity which allows the formation of crystals with sheets of a few hundred particles in hexagonal lattices in the horizontal plane but only a few such sheets stacked upon each other in the vertical direction. The primary interparticle force in the horizontal plane is the shielded Coulomb (Yukuwa) force while in the vertical directions the particles interact in a more complicated manner due to the wake effects of ions passing through the plasma sheath.

**Computer Simulation:** Computer models have been developed in order to better understand such systems[5-7]. In the majority of these, only interactions between near neighbors are considered since the shielded Coulomb interaction decays exponentially. Such simulations typically examine only the bulk of the system by employing periodic boundary conditions. Under these assumptions, the longer-range interactions between particles is negligible due to the decaying exponential of the Coulomb interaction and overall system symmetry.

However, it is becomingly increasingly clear that much of the interesting behavior of dusty plasma crystals may be due to edge effects where the symmetry of the system breaks down. Few, if any, computer simulations have been conducted to examine such finite crystals where the symmetry is reduced.

A finite two-dimensional crystal is modeled here using a Barnes-Hut tree code known as *Box_Tree*[7-9]. *Box_Tree* includes interactions due to gravity in addition to either the shielded or unshielded Coulomb interaction. As can be seen in Figure 1, tree codes divide the system into a set of nested boxes with every box being empty or containing either a single particle or four sub-boxes (eight for three dimensional simulations).

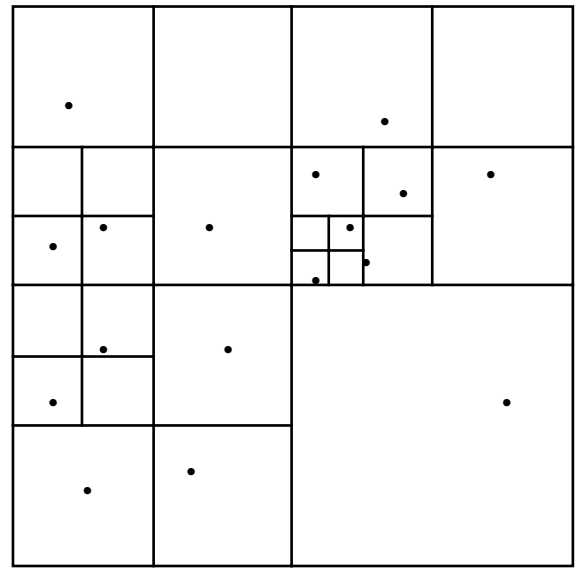

Figure 1. Tree structure.

When calculating the force on a particle the code transverses the tree starting with the largest box (which contains every particle) and subsequently examining smaller and smaller boxes. It subsequently checks each box to determine if the box contains a single particle or sub-boxes and examines the ratio of the size of the box to the distance from the box to the particle. For boxes containing a single particle the force is calculated directly between the particles. If the ratio of the box size to the distance is smaller than a user-defined critical value the force is calculated by examining the multipole moments of the particles in the box. For most near-by particles the force will be calculated directly. For more distant particles the force is typically found using the multi-pole moments of the box. This allows all interparticle interactions to be included while treating near-field interactions with greater accuracy. An additional benefit is that the code scales as $N \cdot \log N$ instead of $N^2$ since most interactions are calculated using the multi-pole moments of collections of particles. The particles are confined by a ring of immobile charged particles, simulating the outer wall of the reference cell.

**Results and Conclusions:** *Box_Tree* has been used to study a number of different configurations. Both the charges on the particles are varied as well as the charge on the containing ring.